# Extremely Narrow, Sharp-Peaked Resonances at the Edge of the Continuum

Ignas Lukosiunas[1*], Lina Grineviciute[2], Julianija Nikitina[2], Darius Gailevicius[1], and Kestutis Staliunas[1,3,4]

[1]Vilnius University, Faculty of Physics, Laser Research Center, Sauletekio Ave. 10, Vilnius, Lithuania
[2]Center for Physical Sciences and Technology, Savanoriu Ave. 231, LT-02300 Vilnius, Lithuania
[3]ICREA, Passeig Lluís Companys 23, 08010, Barcelona, Spain
[4]UPC, Dep. de Fisica, Rambla Sant Nebridi 22, 08222, Terrassa (Barcelona) Spain

*Corresponding author: ignas.lukosiunas@ff.vu.lt

**Abstract:** We report a critical narrowing of resonances of a driven potential well, when their eigenfrequencies approach the edge of the continuum. The resonances also obtain unusual sharp-peak shapes at the continuum boundary. The situation can be realized for the electromagnetic wave propagating across the dielectric thin films with a periodically modulated interface(s). We show the general phenomenon semi-analytically on a simplified model of a driven quantum potential well, also by rigorous numerical analysis of Maxwell equations for the wave propagation across the thin film with a modulated interface(s). We justify the phenomenon experimentally, by the measurements of light reflection from the dielectric thin film deposited on a periodically modulated surface. The narrow and sharp-peak resonances can be used for an efficient narrow-band frequency- and spatial filtering of light.

**Introduction:** It is well known that the wavefunctions in a potential well can form bound states for their energies below the background of the potential well, and the continuum states for their energies above the background. Although the wave behavior in the potential wells has been established since the beginning of quantum mechanics [1], new findings are being reported. One such example is the phenomenon of Bound States in the Continuum (BIC), recently proposed for the potentials of a special form [2,3]. Here we report another effect never considered before – critical narrowing and extreme sharpening of the resonances of bound states when their energies approach the background energy of the potential well, just before crossing the boundary of the continuum.

The resonances of the discrete states follow the universal Lorentzian shapes when the energies are deep in the potential well. What happens when the potential is continuously deformed so, that the energy of one of its discreet states crosses the Edge of the Continuum (EOC), as illustrated in Fig.1 The letter shows that the width of the resonances strongly decreases, and the resonances obtain sharp-peaked shapes at the crossing point. Moreover, the scaling of the resonance width with the coupling constant also becomes unusual: whereas the width of the Lorentz resonance scales as the power of 2 with the coupling constant, the width of the resonance at the EOC scales as a power of 4. In this way, the resonances with special properties are reported, which on the one hand, are intriguing mathematical-physical objects, on the other hand, may have huge practical application potential. The effect, in particular, might be useful to realize extremely narrow band and sharp-contrast frequency- and angular (spatial) filters.

In this letter we mathematically relate the problem of electromagnetic wave diffraction on interface-modulated thin dielectric films with the problem of a particle's wavefunction in a driven potential well, Fig.1. The wave propagation across such thin films has been previously studied, mostly numerically, by Rigorous Coupled Wave approach, in the context of its angle-wavelength transmission peculiarities [4-7], or Fano-like resonances [8-11]. The analogy between the thin films and the driven potential well, established in the present work, allows calculating the energy states in the potential well, corresponding to the thin film planar modes, and exploring these resonances semi-analytically. We calculate the shapes and estimate the width of the resonances in a simplified model system of a driven potential well. We apply rigorous numerical methods to calculate the light reflection/transmission through the modulated thin films and identify these narrow resonances in reflection angle-wavelength spectra. Finally, we fabricate a structure by Physical Vapor Deposition (PVD) on a corrugated surface and measure the predicted narrow-band sharp resonances in the reflection angle-wavelength spectra.

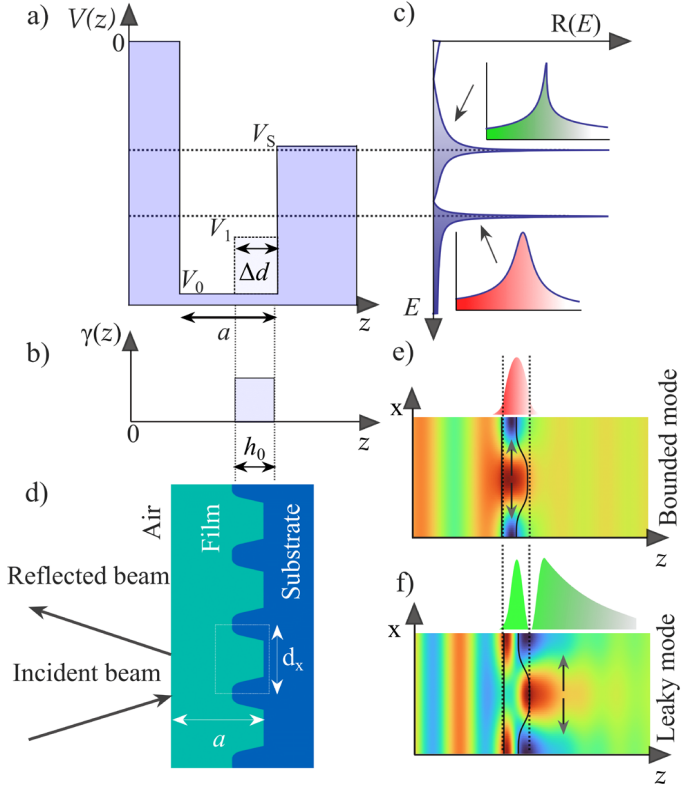

**FIG. 1**. Driven potential well (a,b) is equivalent to the film of high refraction index with periodically modulated interfaces (d). The interface modulation couples the nearly-normal FP mode into the waveguided modes (e) or leaky surface modes (f), resulting in usual Lorenz or sharp-peaked resonances, respectively (c). The modification of the refraction index in the coupling area is shown by the line in (a).

**Simplified model.** In analytical treatment, we separate the electromagnetic radiation into the near-to-normal part (corresponding to the incident/ reflected/ transmitted modes) and the guided radiation part (parallel to the film surface). Specifically, we define the near-to-normal wave as: $A_0(x,z,t) = A_0(z)exp(ik_{0,z}z + ik_{0,x}x - i\omega t)$, here $k_{0,x}^2 + k_{0,z}^2 = k_0^2 = \omega^2/c^2$. This can be considered as the Fabry-Perrot (FP) mode of the resonator formed by a thin film. We also define the guided radiation as: $A_1(x,z,t) = A_1(z)exp(ik_{1,x}x - i\omega t)$. The near-to-normal and guided components are mutually coupled due to the periodic modulation of the thin film interface, with the period $d_x$ of the order of the wavelength of the incident wave: $k_{1,x} = k_{0,x} \pm q$. We consider, for analytical treatment, only the first diffraction order: $q = 2\pi/d_x$, also only one diffracted mode, for instance, the one with $k_{1,x} = k_{0,x} + q$. Then the stationary wave equation for the guided radiation component $A_1(z)$ reads:

$$\frac{\partial^2 A_1(z)}{\partial z^2} + \left(\frac{\varepsilon(z)\omega^2}{c^2} - (k_{0,x} + q)^2\right) A_1 = 0 \quad (1)$$

The equation (1) is rigorous for the case of *s*-polarized waves, when the electric field oscillates along the modulation grooves, i.e., along the coordinate *y*. (The *p*-polarized waves also allow a simplified treatment, see the Suppl.B, resulting in the same qualitative physical picture.) Eq (1) is identical to the stationary Schrödinger equation for the quantum particle with the energy $E = k_0^2 - (k_{0,x} + q)^2$ in the potential well with the spatial profile $V(z) = (1 - \varepsilon(z))k_0^2$ as illustrated in Fig.1.a.

Such a potential supports the bound states, which corresponds to the planar modes of the thin film. Coupling between the FP radiation $A_0(z)$ to these planar modes $A_1(z)$ results in the Fano-type resonances, studied, for instance, in [8-11]. We introduce the coupling between FP radiation and the guided radiation by additional external driving for the latter component:

$$\frac{\partial^2 A_1(z)}{\partial z^2} + (E - V(z))A_1(z) - i\gamma(z)e^{ik_{0,z}z}A_0(z) = 0 \quad (2)$$

$\gamma(z)$ is the normalized coupling profile along the z. see Fig.1.b. For analytical treatment, we consider the uniform coupling over the area in z of thickness $h_0$, $\gamma(z) = \gamma_0/h_0$, $\gamma_0$ being a dimensionless net coupling coefficient, and $\gamma(z) = 0$ elsewhere. The estimation of $\gamma_0$ in the limit of a shallow harmonic modulation of the interfaces has been provided, for instance, in [11]: $\gamma_0 \approx h_0 \Delta n/\lambda$, where $h_0$ is the amplitude of the surface modulation of the film: $h(x) = h_0/2 \cos(qx)$, and $\Delta n$ is the difference in the refraction indexes of materials forming the modulated interface. To maximally fit the quantum-potential based model to the experiment we also modify the potential $V(z)$ by deforming its bottom on the interaction section: $V_1 = fV_0 + (1-f)V_s$, i.e., considering f-weighted average of the bottom and right-background potentials, $V_0$ and $V_s$. The presence of a step, indicated by the line in Fig.1.a is not essential to the main results.

Eq. (2) can be solved by finding piecewise the wavefunction in each sector of the potential, *s*, along the z: $A_{1,s}(z) = a_{0,s}e^{ik_{0,z}z} + a_{+,s}e^{ik_{z,s}z} + a_{-,s}e^{-ik_{z,s}z}$ with $k_{z,s}^2 = (E - V_s)$, $a_{0,s} = i\gamma_s A_{0,s}/(E - V_s - k_0^2)$, and matching the solutions and their derivatives on the interfaces between the sectors to determine the $a_{+,s}$ and $a_{-,s}$. The solutions were obtained in an explicit analytical form for the driven potential shown in Fig.1; however, they are not analytically tractable due to their algebraic complexity.

To calculate the transmission/reflection of the incident field through a film, it is convenient to introduce the gain of the FP mode (see Suppl.C):

$$g_0 = \frac{i}{A_0} \int \gamma(z) A_1(z) e^{-ik_{0,z}z} dz \quad (3)$$

The physical sense of gain $g_0$ is the feedback from the guided modes back to the FP mode. The transmission of the incident wave through the thin film then reads: $t_{FP} = (1 - g_0/t^2)^{-1}$. (We consider that the transmission coefficient through the interfaces, $t$, is equal for both interfaces.) The reflection of the incident radiation from the film follows from the relation: $T + R = |t_{FP}|^2 + |r_{FP}|^2 = 1$.

The analysis above model identifies the Fano-type resonances [12] following universal Lorentzian shapes. We varied the depth of the potential well $V_0$ to tune the eigenfrequencies of the resonances to explore the crossing of the highest energy resonance through the EOC. The width of the resonance remains nearly constant when its frequency remains deep in the potential well, but starts decreasing, when approaching the EOC. Meanwhile, the maximum value of the reflection coefficient remains unity. The reflection coefficient decreases after crossing the continuum boundary, as indicated in Fig.2.a. The shape of the resonance curve becomes sharp-peaked at EOC, in contrast to the universal Lorentz shape.

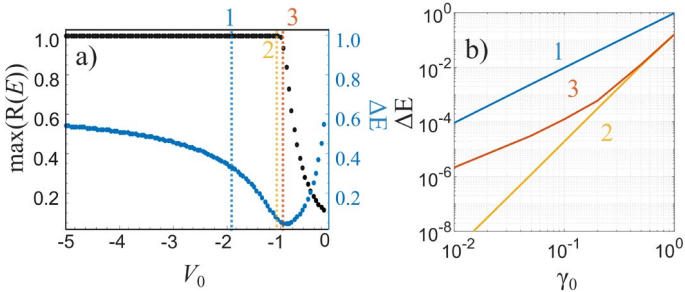

**FIG.2.** a) The maximum of reflection coefficient from the potential well, $max(R(E))$, and the half-width of the reflection resonance $\Delta E_0$ depending on $V_0$. $\gamma_0 = 0.75$. b) The half-width of the reflection resonance $\Delta E_0$ depending on the coupling constant $\gamma_0$ at the EOC, $V_0 = -1$, yellow line, slightly below the EOC, $V_0 = -1.05$, red line, and deep below the EOC, $V_0 = -2$, blue line. Other parameters: $a = \pi/2$, $h_0 = \pi/10$.

**The resonances:** The driven potential well, Fig.1, did not lead to simple algebraic expressions. In order to maximally simplify the problem and to get analytic estimations, we further reduced the model. We considered a semi-infinite potential well, with $V \to \infty$ for $z < 0$, and constant driving force throughout the entire bottom of the potential well $0 < z < h_0$. The analysis (see Suppl.C) lead to the tractable expressions of the gain function in a limiting case of the narrow and deep potential well ($ak_{0,z} \ll 1$, $|V_0| \gg 1$):

$$g_0 = \frac{\gamma_0^2}{a^2} \frac{i\sqrt{V_0}}{k_1^3 k_2} \frac{2k_2^2 + (k_1^2 - 2k_2^2)cos(ak_1)}{cos(ak_1 + \phi)} \quad (4)$$

here $tan(\phi) = ik_2/k_1$, $k_1 = \sqrt{E - V_0}$ and $k_2 = \sqrt{E}$ are the wavenumbers of the wavefunction in the potential well and outside the well, respectively. For the bounded states $E < 0$ the $k_2$ is imaginary valued, whereas $k_1$ remains real-valued.

The poles of (4), $cos(ak_1 + \phi) \to 0$, indicate the resonances, which occur for $(ak_1 + \phi) \to \pi(2n + 1)/2$, where the integer $n$ counts the resonances starting from $n = 0$. We simplify (4) in two different asymptotics:

1) the lowest energy-bound state lies deep in the potential well, $E, V_0 \to -\infty$. Then $|k_1| \ll |k_2|$, $\phi \to -\pi/2$, $cos(ak_1 + \phi) \to 0$, $ak_1 \to \pi$ (for the lowest energy bound state):

$$g_0 = \frac{2i\gamma_0^2}{a^2} \frac{1}{\sqrt{a^2 - \pi^2/V_0}} \frac{1}{\Delta E} \quad (5)$$

In the limit $V_0 \to -\infty$, the half-width of the gain line simplifies to: $\Delta E_0 = 2\gamma_0^2/a^3$, and scales with the coupling coefficient as $\Delta E_0 \sim \gamma_0^2$. This case corresponds to the Fano-like resonances of the discrete states.

2) The lowest energy bound state coincides with the background level on the right side of the potential well at $z > a$, $E = 0$. Then $|k_2| \ll |k_1|$, $\phi \to 0$, $cos(ak_1) \to 0$, $ak_1 \to \pi/2$, and we obtain:

$$g_0 = \frac{-8\gamma_0^2}{\pi a^2} \frac{1}{\sqrt{\Delta E}} \quad (6)$$

The half-width of the gain line becomes: $\Delta E_0 = 64\gamma_0^4/(\pi^2 a^4)$, and scales with the coupling coefficient as $\Delta E_0 \sim \gamma_0^4$.

The (6) is the main result of our analytical study. For the positive detuning $\Delta E > 0$, i.e., within the continuum, the gain is real-valued, whereas for negative detuning $\Delta E < 0$, i.e., below the EOC, the gain is imaginary. This has consequences on the reflection coefficient from the thin film. For both positive and negative $\Delta E$, the reflection coefficient asymptotically decreases as $R_{FP} \approx 1 - |\Delta E|/\Delta E_R$, with $\Delta E_R = 64\gamma_0^4/(\pi^2 t^4 a^4)$, which results in a sharp-peaked shape of the resonance.

These analytic predictions from the simplest model were checked by numerically calculating the driven potential model, shown in Fig.1. The sharp-peaked resonances at the edge of the continuum were identified. The calculated width of the resonances (Fig.2), indeed, shows the predicted scaling. We recover the standard scaling of

Lorenz resonances $\Delta E_0 \sim \gamma_0^2$ for the energies deep below the EOC, and justify the $\Delta E_0 \sim \gamma_0^4$ scaling at the EOC.

**Full model.** To prove the validity of the results obtained on simplified models, we performed the analysis using the rigorous coupled wave method [13,14], i.e., solving the Maxwell equations without any approximations. Our own solver was developed for that purpose [15]. The results are summarized in Fig. 3.

The full model accounts for both (right or left propagating) guided modes. Therefore in Fig.3 the pattern of the left- and right-inclined resonances in parameter space of incidence angle, and the wavelength $(\theta, \lambda)$ is observed. The physical parameter space of the thin film $(\theta, \lambda)$ relates with the parameter space of the potential $(V_0, E)$ via above-presented relations: $E = k_0^2 - (k_0 sin(\theta) + q)^2$ , $V_0 = (1 - \varepsilon_{film})k_0^2$ . The EOL corresponds to a particular set of parameters in $(V_0, E)$, and equivalently in $(\theta, \lambda)$. These particular points are indicated in Fig.3.a by arrows for left and right-guided waves. The inset shows the sharp-peaked asymmetric resonance by crossing the EOL point along a particular $\theta = const$.

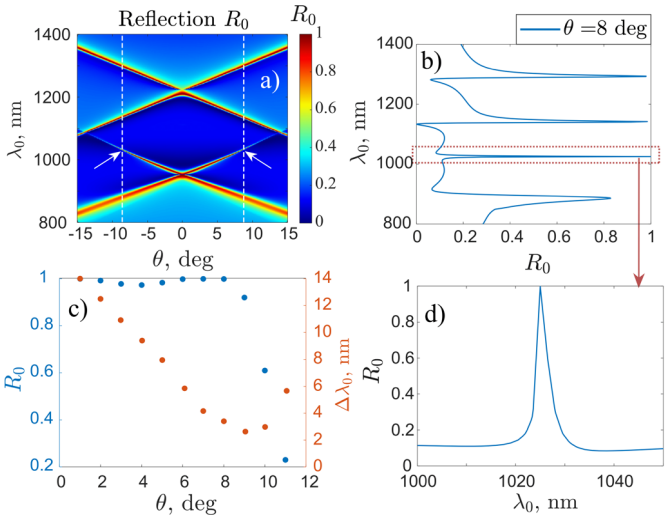

**FIG.3.** *The map of the reflection coefficient in the plane of angle-wavelength $(\theta, \lambda)$ (a), and its cross-section at a given angle $\theta$ (b). The arrows indicate the points corresponding to the resonance at the EOC. The top reflection coefficient and the resonance width of the highest frequency- smallest wavelength mode, depending on $\theta$ (c). The zoom of the reflection profile around the continuum edge (d). The calculations were performed by RCWA with the parameters: modulation depth $h_0 = 200$ nm, modulation period $d_x = 625$ nm, layer thickness $d_z = 460$ nm., refraction indices $n_{film} = 2.25$, $n_{substr.} = 1.5$.*

Resonance lines with only the smallest wavelength (highest energy) cross the EOC in Fig.3. The other resonances, with the larger wavelength (lower energies), remain deep in the potential well. Consequently, their resonances do not develop sharp-peaks, and remain Lorentzian, as Fig.3.b. indicates.

The dependence of the top reflection coefficient and the resonance's width on the angle is shown in Fig.3.c, in accordance with the conclusions from the simplified treatment above.

Finally, the scaling laws of the resonance width with the coupling coefficient (depth of the film interface modulation) were checked. If the usual Lorenz resonances show the well-established power 2 law, the resonances at the edge follow the power 4 law, as the Fig.4.b evidences.

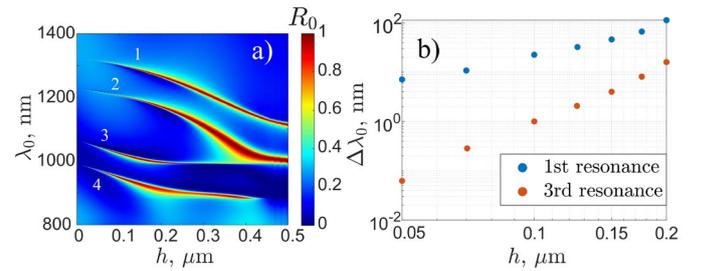

*Fig. 4.(a) The map of the reflection coefficient in the plane of modulation depth vs wavelength shows. The width of the resonances versus modulation depth is shown in (b). The parameters are same as in Fig.3.*

**Experimental realization.** For fabricating such a structure, we used a commercially available fused silica surface grating (modulation period 625 nm, modulation depth 220 nm) with nearly sinusoidal surface (see Fig.5. a and b). The grating was used as a substrate for thin film deposition by PVD [16]. The ion beam sputtering technology [17] was used for $Nb_2O_5$ layer fabrication (thickness = 530 nm, refractive index – 2.23); see also [18] for technological details. The profile of the thin film surface remained almost the same as that of the substrate modulation (see fig.5.b).

Spectrophotometric measurements recorded reflection maps for the fabricated sample. Linearly polarized light was used for two perpendicular polarizations: s- and p-, where s-polarization is parallel to the grating lines on the sample. The angle between the normal of the grating and the detector was varied from 0° to 15° by steps of 0.5°. The resulting reflection maps are presented in Figs. 5.c, and 5.e.

The measured reflection maps of the sample correspond well to those following from the RCWA simulations. The lines of the Fano resonances were observed, narrowing and terminating at EOC, and indicating the sharp-peaked shapes, see Fig.5 (c-f).

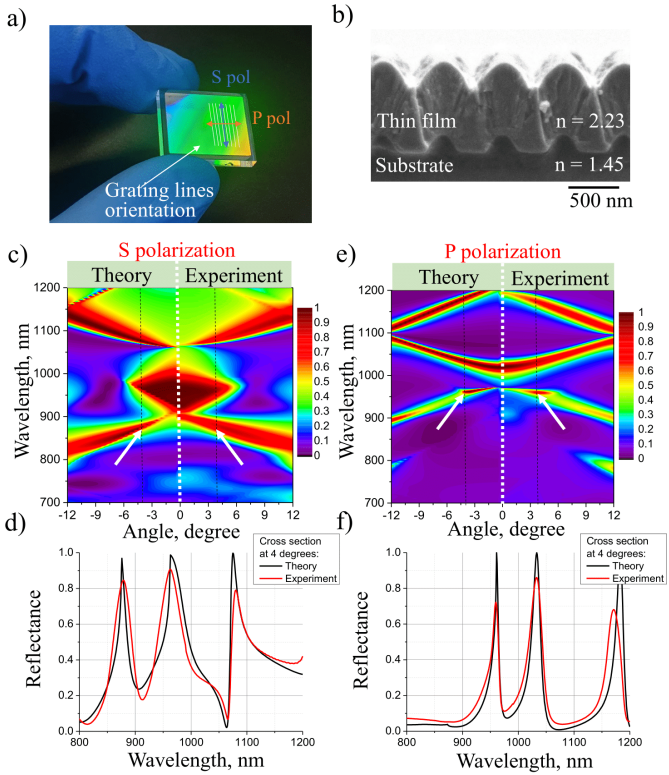

**FIG.5.** *Experimental results, compared with RCWA results: Photo of fabricated samples (arrows indicate polarization directions (a), and the SEM image of specimen sample cross-section (b). Numerical/experimental maps of the reflection in the plane of angle-wavelength for (c,d) S polarization and (e,f) P polarization, with their cross-sections at 4 degrees of light incidence angle. The arrows indicate the points where the left and guided mode crosses continuum boundary.*

Remarkably, the sharp-peaked resonances were indicated not only for the S-polarization, as predicted by the above analysis, but also for P-polarization. Also, the both interfaces of the thin film were modulated. Current technical limitations of thin film fabrication did not allow to obtain the modulation of only one interface, which strictly correspond to the analytically studied cases. This indicates that the predicted and observed effect is generic (not restricted to the Schrödinger equation model), and robust with respect to nonessential modifications of the geometry of driven potential.

**In summary,** we predicted a new general effect, a critical narrowing sharpening of the peak of the resonances when their eigenfrequencies approach the EOC. Importantly, the maximum reflection coefficient (at the middle of the resonance) remains unity. We realized the effect for the wave propagation/reflection through the thin films with periodically modulated interfaces, and we demonstrated the predicted effect measuring reflection coefficients depending on the frequency and incidence angle.

For the usual Lorentz resonances, the gain function $g_0$ is imaginary-valued and symmetric, which gives a smooth variation of the phase delay of the response from 0 to $\pi$, with a value $\pi/2$ at the resonance, as well as smooth-peak of the resonance curve. In reported case the gain function $g_0$ changes from imaginary to real at EOC, which gives an abrupt $\pi/2$ jump of phase delay, as well as an unusual sharp-peaked shape of the resonance.

To demonstrate the essence of the effect, we intended to isolate the points corresponding to the EOC, which allows us to interpret the phenomena in frames of a maximally simple model of only one guided wave. Additional effects could be expected at the coalescence of such two exceptional points corresponding to left and right-guided waves. The crossing resonance lines in parameter space $(\theta, \lambda)$, corresponding to left-right guided waves, can form in this plane sharp-edged triangles, which could be used for efficient spatial filtering. For instance, such spatial filtering effects were realized in [19,20] using the crossings of the usual, symmetric smooth-top resonance lines of Lorentzian form. Here, the asymmetries and sharp features of the coalescing resonances at the edge of the continuum can lead to advanced spatial filters with exotic characteristics, like sharp edges and flat tops, among others.

**Acknowledgments:** This work has received funding from the Spanish Ministerio de Ciencia e Innovación under grant No.385 (PID2019-109175GB-C21), European Social Fund (project No 09.3.3- LMT-K712-17- 0016) under grant agreement with the Research Council of Lithuania (LMTLT), also from Horizon 2020 ERA.NET-COFUND program project MiLaCo (Project No. S-M-ERA.NET-20-2) under grant agreement with the Research Council of Lithuania (LMTLT).